The effects of energy and commodity prices on commodity output in a three-factor, two-good general equilibrium trade model[1]

Authors: Yoshiaki Nakada

Division of Natural Resource Economics, Faculty of Agriculture, Kyoto University, Kita-Shirakawa-Oiwake-cho, Sakyo, Kyoto, Japan 606-8502

Abstract:

We analyze the effects of energy and commodity prices on commodity output using a three-factor, two-good general equilibrium trade model with three factors: capital, labor, and imported energy. We derive a sufficient condition for each sign pattern of each relationship to hold, which no other studies have derived. We assume factor-intensity ranking is constant and use the EWS (economy-wide substitution)-ratio vector and the Hadamard product in the analysis. The results reveal that the position of the EWS-ratio vector determines the relationships. Specifically, the strengthening (resp. reduction) of import restrictions can increase (resp. decrease) the commodity output of exportables, if capital and labor, domestic factors, are economy-wide complements. This seems paradoxical.

1. Introduction

Thompson (1983) conducted a study on functional relations in a three-factor, two-good neoclassical model (hereinafter, 3 x 2 model), where one factor is internationally mobile. Thompson called it the 3 x 2 mobile factor model. For example, using eq. (11), he stated that "rising (falling) payment to the mobile factor cannot increase (decrease) both outputs" (p. 47)."[2] Thompson (1983,

---

[1] An earlier version was titled, "The energy price -- commodity output relationship and the commodity price -- commodity output relationship in a three-factor, two-good general equilibrium trade model with imported energy," uploaded by Yoshiaki Nakada in Nov. 2017. Available at https://www.researchgate.net/publication/321347535_The_energy_price_-_commodity_output_relationship_and_the_commodity_price_-_commodity_output_relationship_in_a_three-factor_two-good_general_equilibrium_trade_model_with_imported_energy.

[2] However, I am uncertain whether Thompson's (1983, pp. 47-8) proof is plausible. Specifically, he used $D'(<0)$ to prove the property of eq. (11), where $D'$ is "the system determinant of the usual 3 x 2 model with all factor payments endogenous". However, we do not need this for the proof.

p.48) continued, "With $p_1$ rising, for instance, $x_1$ rises and $x_2$ falls." He asserted that if the price of good 1 increases, the output of good 1 rises and that of good 2 falls. However, Thompson did not provide proof supporting this statement.

Thompson (1994) analyzed the 3 x 2 model using three factors (capital, labor, and imported energy), and called it the 3 x 2 international energy model. According to Thompson (1994, pp. 198-9), there are three sign patterns of the energy tariff–commodity output relationship. However, he did not show a sufficient condition for each sign pattern of that relationship to hold. Thompson (2000) analyzed the same model to explore the effect of energy tax on wage.

Thompson (2014) analyzed a 3 x 2 model with three factors (capital, labor, and imported energy) in elasticity terms, whereas Thompson (1983, 1994, 2000) examined in differential forms. Thompson's (2014) model explicitly included energy tariff and (national) income including tariff revenue.[3] Thompson (2014, p. 65) suggested that factor intensity and substitution were important for his analysis.

In summary, Thompson (1983, 1994, 2000, 2014) uses the 3 x 2 model, where one factor payment is exogenous. I am uncertain whether all of Thompson's results are plausible. For example, the equation in Thompson (2014) denoting the commodity price–energy import relationship includes an error (see the Appendix A).

Accordingly, the following questions emerge:

(i) Can we derive a sufficient condition for each sign pattern of the energy price–commodity output relationship to hold?

(ii) Is Thompson's (1983) statement—if the price of good 1 increases, then the output of good 1 rises and that of good 2 falls—plausible?

To the best of our knowledge, no study except Thompson (1983, 1994, 2000, 2014) has analyzed a 3 x 2 model in which one factor payment is exogenous.[4]

---

Originally, he used it to prove the property of eq. (9). I do not discuss this further as it is beyond the scope of this paper.

[3] For example, Thompson (2014) analyzed the effects of energy tariffs on energy import, commodity output, and domestic factor price. In addition, he examined the effects of capital, labor, and commodity price on national income. He then conducted a simulation study under the assumption of the Cobb–Douglas production functions.

[4] Takeda (2005, pp. 1-2) adopted a model with two goods, where each sector employs three factors (capital, labor, and emission tax, which is sector specific). Hence, his model is a four-factor two-good trade model, where two factor payments are specific and exogenous. However, Takeda (2005, p. 1) stated, "Thus, the model has a structure similar to the standard 2 × 3 [3 x 2 in our expression] HO [or Heckscher-Ohlin] model employed in Batra and Casas (1976) and Jones and Easton (1983)." Furthermore, Ishikawa and Kiyono (2006, p. 432, note 3) stated, "Takeda (2005) had extended our model [2 x 2 x 2 model] to a two-good, three-factor model." On the other hand, Heutel and Fullerton

In this study, we conduct a detailed analysis on functional relations in a 3 x 2 model with three factors (capital, labor, and imported energy). We do not include energy tariffs for ease of analysis. The objective of this study is two-fold. First, we derive a sufficient condition for each sign pattern of the energy price–commodity output relationship to hold. Next, we analyze the commodity price–commodity output relationship.

Batra and Casas (1976) (hereinafter BC) published an article on functional relations in a 3 x 2 model with all factor payments endogenous. By moving some terms in the basic equations for this model, we derive those for a 3 x 2 model with imported energy. Nakada (2017) defined the EWS-ratio vector based on 'economy-wide substitution' (hereinafter EWS) originally defined by Jones and Easton (1983) (hereinafter JE) and applied it in the analysis of the 3 x 2 model of BC's original type. We also use the EWS-ratio vector in the present study.

We assume factor-intensity ranking is constant. Similar to Thompson (1994, 2000), we assume that sector 1 is relatively energy-intensive, sector 2 is relatively capital-intensive, labor is the middle factor, and energy and capital are extreme factors. We also assume factor-intensity ranking for the middle factor is constant.[5] That is, we assume the middle factor is used relatively intensively in sector 1. We conduct the analysis in elasticity terms.

The remainder of this paper is organized as follows. Section 2 presents the model. In section 2.1, we explain the basic structure of the model. We make a system of linear equations using a 5 x 5 matrix. In section 2.2, we assume factor-intensity ranking. In section 2.3, we define the EWS-ratio vector based on EWS for the analysis. We derive the important relationship among EWS-ratios and draw the EWS-ratio vector boundary, which is useful for our analysis. In section 2.4, we derive the solutions for the system of linear equations. In section 2.5, we analyze the energy price–commodity output relationship using EWS-ratios. In section 2.6, we examine the commodity price–commodity output relationship. In section 2.7, we derive the factor endowment–commodity output relationship. Section 3 concludes the paper. Appendix A presents Thompson's (2014) equations that include errors. Appendix B derives the solution of eq. (19). Appendix C shows the expansion of eq. (51). Appendix D derives the solutions of eq. (51). Section 2.3 contains the similar content as Nakada (2017).

We classify other studies analyzing the model with one factor payment exogenous as follows.
(i)     Studies that analyzed a two-factor, two-good, two-country model (or 2 x 2 x 2 model), where capital is internationally mobile. For example, Mundell (1957, p. 322), Kemp (1966), Jones

---

(2010, pp. 3-4) applied a 3 x 2 model, in which pollution tax is specific to the dirty sector, or polluting sector. In their model, commodity price is endogenous. Hence, this model is not a 3 x 2 trade model (for details of their model, see Fullerton and Heutel (2007, pp. 574-5)).

[5] Thompson (2014, eq. (9)) assumed factor-intensity ranking differently. He considered that sector 1 is relatively energy-intensive and sector 2 is relatively labor-intensive. In addition, Thompson (1983, 1994, 2000, 2014) did not assume the factor-intensity ranking for the middle factor.

(1967), Chipman (1971), Jones and Ruffin (1975), and Ferguson (1978, p. 374).

(ii) Studies that analyzed a 2 x 2 x 2 model, where each sector employs domestic labor and greenhouse gas emission traded internationally. For example, Ishikawa and Kiyono (2006, p. 432-3), and Ishikawa, Kiyono, and Yomogida (2012, p. 187-8).

(iii) Studies that analyzed the simplest type of a 3 x 2 model, what you call, specific factors model, where the factor specific to one sector is internationally mobile. For example, Srinivasan (1983, p. 292), Brecher and Findlay (1983), Thompson (1985), Dei (1985), Bandyopadhyay and Bandyopadhyay (1989), and Ogino (1990).

(iv) Studies that analyzed a N-factor, M-good model, where some factors are internationally mobile. For example, Svensson (1984) and Ethier and Svensson (1986, p. 22). Ethier and Svensson (1986) examined some theorems such as the Rybczynski and Stolper-Samuelson theorems.

2. Model

2.1. Basic structure of the model

We assume as follows. Products and factor markets are perfectly competitive. The supply of all factors is perfectly inelastic. Production functions are homogeneous of degree one and strictly quasi-concave. All factors are not specific and perfectly mobile between sectors, and factor prices are perfectly flexible. These two assumptions ensure the full employment of all resources. The country is small and faces exogenously given world prices, or the movement in the price of a commodity is exogenously determined. The movements in factor endowments are exogenously determined.

The 3 x 2 model in the present study uses the same symbols as in Nakada (2017) for the basic equations. However, in this study, $w_T$ and $V_T$ denote energy price and the amount of imported energy, which are exogenous and endogenous variables.

Full employment of factors implies

$$\sum_j a_{ij} X_j = V_i \quad i = T, K, L \tag{1}$$

where $X_j$ denotes the amount produced of good $j$ ($j = 1, 2$); $a_{ij}$ denotes the requirement of input $i$ per unit of output of good $j$ (or the input-output coefficient); $V_i$ denotes the supply of factor $i$; $T$ is the energy, $K$ capital, and $L$ labor.

In a perfectly competitive economy, the unit cost of production of each good must just equal its price. Hence,

$$\sum_i a_{ij} w_i = p_j, \ j = 1, 2, \tag{2}$$

where $p_j$ is the price of good $j$, and $w_i$ is the reward of factor $i$.

BC (p. 23) stated, 'With quasi-concave and linearly homogeneous production functions, each input-output coefficient is independent of the scale of output and is a function solely of input prices:'

$$a_{ij} = a_{ij}(w_i), \ i = T, K, L, \ j = 1, 2. \tag{3}$$

The authors continue, 'In particular, each $C_{ij}$ [$a_{ij}$ in our expression] is homogeneous of degree zero in all input prices.'[6]

Equations (1)-(3) describe the production side of the model. These are equivalent to eqs (1)-(5) in BC. The set includes 11 equations in 11 endogenous variables ($X_j$, $a_{ij}$, $w_K$, $w_L$, and $V_T$) and five exogenous variables ($V_K, V_L, w_T$, and $p_j$). The small country assumption simplifies the demand side of the economy. Totally differentiate eq. (1):

$$\Sigma_j (\lambda_{ij} a_{ij}^* + \lambda_{ij} X_j^*) = V_i^*, \ i = T, K, L, \tag{4}$$

where an asterisk denotes a rate of change (e.g., $X_j^* = dX_j / X_j$), and where $\lambda_{ij}$ is the proportion of the total supply of factor $i$ in sector $j$ (that is, $\lambda_{ij} = a_{ij} X_j / V_i$). Note that $\Sigma_j \lambda_{ij} = 1$.

The minimum unit cost equilibrium condition in each sector implies $\Sigma_i w_i da_{ij} = 0$. Hence, we derive (see JE (p. 73), BC (p. 24, note 5)),

$$\Sigma_i \theta_{ij} a_{ij}^* = 0, \ j = 1, 2, \tag{5}$$

where $\theta_{ij}$ is the distributive share of factor $i$ in sector $j$ (that is, $\theta_{ij} = a_{ij} w_i / p_j$). Note that $\Sigma_i \theta_{ij} = 1$; $da_{ij}$ is the differential of $a_{ij}$.

---

[6] From the condition of cost minimization, we can show that $a_{ij}$ is homogeneous of degree zero in all input prices (see Samuelson (1953, chapter 4, p. 68), Nakada (2017, eq. 3)).

Totally differentiate eq. (2):

$$\Sigma_i \theta_{ij} w_i^* = p_j^*, \ j=1,\ 2. \tag{6}$$

Totally differentiate eq. (3) to obtain

$$a_{ij}^* = \Sigma_h \varepsilon^{ij}_h w_h^* = 0, \ i=T,\ K,\ L,\ j=1,\ 2, \tag{7}$$

where

$$\varepsilon^{ij}_h = \partial log a_{ij} / \partial log\ w_h = \theta_{hj} \sigma^{ij}_h. \tag{8}$$

$\sigma^{ij}_h$ is the AES (or the Allen-partial elasticities of substitution) between the ith and the hth factors in the jth industry. For an additional definition of these symbols, see Sato and Koizumi (1973, p. 47-9), BC (p. 24). AESs are symmetric in the sense that

$$\sigma^{ij}_h = \sigma^{hj}_i. \tag{9}$$

According to BC (p. 33), 'Given the assumption that production functions are strictly quasi-concave and linearly homogeneous,'

$$\sigma^{ij}_i < 0. \tag{10}$$

Because $a_{ij}$ is homogeneous of degree zero in all input prices, we have

$$\Sigma_h \varepsilon^{ij}_h = \Sigma_h \theta_{hj} \sigma^{ij}_h = 0, \ i=T,\ K,\ L,\ j=1,\ 2. \tag{11}$$

Equations (7) to (11) are equivalent to the expressions in BC (p. 24, n.6). See also JE (p. 74, eqs (12)-(13)).

Substituting eq. (7) in (4), we derive

$$\Sigma_h g_{ih} w_h^* + \Sigma_j \lambda_{ij} X_j^* = V_i^*, \ i=T,\ K,\ L, \tag{12}$$

where

$$g_{ih} = \Sigma_j \lambda_{ij} \varepsilon^{ij}_h, i,\ h=T,\ K,\ L. \tag{13}$$

This is the EWS (or the economy-wide substitution) between factors $i$ and $h$ defined by JE (p. 75).

$g_{ih}$ is the aggregate of $\varepsilon^{ij}{}_h$. JE (p. 75) stated, 'Clearly, the substitution terms in the two industries are always averaged together. With this in mind, we define the term $\sigma^i{}_k$ to denote the economy-wide substitution towards or away from the use of factor $i$ when the $k$th factor becomes more expensive under the assumption that each industry's output is kept constant.'

We can easily show that

$$\Sigma_h g_{ih} = 0, i = T, K, L, \tag{14}$$

$$g_{ih} = (\theta_h / \theta_i) g_{hi}, i, h = T, K, L, \tag{15}$$

where $\theta_i$ and $\theta_j$ are, respectively, the share of factor $i, i = T, K, L$, and good $j, j = 1, 2$ in gross national income. That is, $\theta_j = p_j X_j / I$, $\theta_i = w_i V_i / I$, where $I = \Sigma_j p_j X_j = \Sigma_i w_i V_i$. See BC (p. 25, eq. (16)).[7] Hence, we obtain $\lambda_{ij} = (\theta_j / \theta_i) \theta_{ij}$ (see JE (p. 72, n. 9)). Note that $\Sigma_j \theta_j = 1$, $\Sigma_i \theta_i = 1$. $g_{ih}$ is not symmetric. Specifically, $g_{ih} \neq g_{hi}, i \neq h$ in general. On eq. (15), see also JE (p. 85).

From eqs (8), (10), and (13), we can show that

$$g_{ii} < 0. \tag{16}$$

From eqs (14) and (16), we derive

$$g_{KT} + g_{KL} = -g_{KK} > 0, \quad g_{TK} + g_{TL} = -g_{TT} > 0, \quad g_{LK} + g_{LT} = -g_{LL} > 0. \tag{17}$$

From eqs (15) and (17), we can easily show that

$$(g_{LK}, g_{LT}, g_{KT}) = (+, +, +), (-, +, +), (+, -, +), (+, +, -). \tag{18}$$

At most, one of the EWSs $(g_{LK}, g_{LT}, g_{KT})$ can be negative.

Moving the term of $w_T^*$ in eqs. (6) and (12) to the right-hand side and that of $V_T^*$ in eq. (12) to the left-hand side, make new equations. Next, combine these equations to make a system of linear equations. Using a 5 x 5 matrix, we obtain

$$\mathbf{AX} = \mathbf{P}, \tag{19}$$

---

[7] Note that BC denoted $I$ as national income in a 3 x 2 model with all factor payments as endogenous, whereas Nakada (2017) called $I$ total income. On the other hand, we define national income as $I' = I - w_T V_T$. In addition, Thompson (2014) did not use $\theta_i$ and $\theta_j$.

where $\mathbf{A} = \begin{bmatrix} 0 & \theta_{K1} & \theta_{L1} & 0 & 0 \\ 0 & \theta_{K2} & \theta_{L2} & 0 & 0 \\ -1 & g_{TK} & g_{TL} & \lambda_{T1} & \lambda_{T2} \\ 0 & g_{KK} & g_{KL} & \lambda_{K1} & \lambda_{K2} \\ 0 & g_{LK} & g_{LL} & \lambda_{L1} & \lambda_{L2} \end{bmatrix}$, $\mathbf{X} = \begin{bmatrix} V_T^* \\ w_K^* \\ w_L^* \\ X_1^* \\ X_2^* \end{bmatrix}$, $\mathbf{P} = \begin{bmatrix} p_1^* - \theta_{T1} w_T^* \\ p_2^* - \theta_{T2} w_T^* \\ -g_{TT} w_T^* \\ V_K^* - g_{KT} w_T^* \\ V_L^* - g_{LT} w_T^* \end{bmatrix}$.

**A** is a 5 x 5 coefficient matrix, and **X, P** are column vectors.

2.2. Factor-intensity ranking

In this article, we assume

$$\theta_{T1}/\theta_{T2} > \theta_{L1}/\theta_{L2} > \theta_{K1}/\theta_{K2}, \tag{20}$$

$$\theta_{L1}/\theta_{L2}. \tag{21}$$

Eq. (20) is, what you call, 'the factor-intensity ranking' (see JE (p. 69), see also BC (p. 26-7), Suzuki (1983, p. 142)).[8] This implies that sector 1 is relatively energy-intensive, and sector 2 is relatively capital-intensive, labor is the middle factor, and energy and capital are extreme factors (see also Ruffin (1981, p. 180)). Eq. (21) is 'the factor-intensity ranking for middle factor' (see JE (p. 70)). It implies that the middle factor is used relatively intensively in sector 1.

Define that

$$(A, B, E) = (\theta_{T1} - \theta_{T2}, \theta_{K1} - \theta_{K2}, \theta_{L1} - \theta_{L2}). \tag{22}$$

This is the inter-sectoral difference in the distributional share. Using eq. (5), we derive

$$A + B + E = 0. \tag{23}$$

Because we assume eqs (20) and (21) hold, we derive

$$(A, B, E) = (+, -, +). \tag{24}$$

2.3. EWS-ratio vector boundary

In this section, we show the important relationship between EWS-ratios, and we draw the EWS-ratio vector boundary in the figure. This is useful for our analysis.

---

[8] For the details of the factor-intensity ranking, see Nakada (2017).

Each $a_{ij}$ function is homogeneous of degree zero in all input prices (see eq. (3)). From eq. (10), $\sigma^{ij}_i < 0$. This implies (see eq. (39) in Nakada (2017))

$$U' > -\frac{\theta_L}{\theta_K}\frac{S'}{S'+1}, \text{ if } T > 0; U' < -\frac{\theta_L}{\theta_K}\frac{S'}{S'+1}, \text{ if } T < 0, \quad (25)$$

where

$$(S', U') = (S/T, U/T) = (g_{LK}/g_{LT}, g_{KT}/g_{LT}), \quad (26)$$

$$(S, T, U) = (g_{LK}, g_{LT}, g_{KT}). \quad (27)$$

We call $(S', U')$ the EWS-ratio vector. $S'$ denotes the relative magnitude of EWS between factors $L$ and $K$ compared to EWS between factors $L$ and $T$. $U'$ denotes the relative magnitude of EWS between factors $K$ and $T$ compared to EWS between factors $L$ and $T$.

Transform

$$U' = -\frac{\theta_L}{\theta_K}\frac{S'}{S'+1} = -\frac{\theta_L}{\theta_K} + \frac{\theta_L}{\theta_K}\frac{1}{S'+1}, \quad (28)$$

which expresses the rectangular hyperbola. We call this the equation for the EWS-ratio vector boundary. It passes on the origin of O (0, 0). The asymptotic lines are $S' = -1$, $U' = -\theta_L/\theta_K$. We can draw this boundary in the figure (see Fig. 1). $S'$ is written along the horizontal axis and $U'$ along the vertical axis. The EWS-ratio vector boundary demarcates the boundary of the region for the EWS-ratio vector. This implies that the EWS-ratio vector is not so arbitrary, but exists within these bounds.

The sign pattern of the EWS-ratio vector is, in each quadrant (on this, see eq. (18)):

Quad. I: $(S', U') = (+, +) \leftrightarrow (g_{LK}, g_{LT}, g_{KT}) = (+, +, +)$;
Quad. II: $(S', U') = (-, +) \leftrightarrow (g_{LK}, g_{LT}, g_{KT}) = (-, +, +)$;
Quad. III: $(S', U') = (-, -) \leftrightarrow (g_{LK}, g_{LT}, g_{KT}) = (+, -, +)$;
Quad. IV: $(S', U') = (+, -) \leftrightarrow (g_{LK}, g_{LT}, g_{KT}) = (+, +, -)$. (29)

Hence, one of the EWS can be negative at most. We define (for $i \neq h$) that factors i and h are economy-wide substitutes (resp. complements), if $g_{ih} > 0$ (resp. $g_{ih} < 0$).

2.4. Solution

From eqs. (B9) and (B10), we have (see Appendix B)

$$\begin{bmatrix} X_1^*/w_T^* \\ X_2^*/w_T^* \end{bmatrix} = \frac{-1}{\Delta} \begin{bmatrix} C_{T1} \\ -C_{T2} \end{bmatrix}, \qquad (30)$$

$$\begin{bmatrix} X_1^*/p_1^* & X_1^*/p_2^* \\ X_2^*/p_1^* & X_2^*/p_2^* \end{bmatrix} = \frac{-1}{\Delta} \begin{bmatrix} C_{11} & -C_{21} \\ -C_{12} & C_{22} \end{bmatrix}. \qquad (31)$$

Equations (30) and (31) express the energy price–commodity output relationship and the commodity price–commodity output relationship, respectively.

2.5. Energy price–commodity output relationship

In this subsection, we analyze the energy price–commodity output relationship. Using eq. (15), eliminate $g_{KL}$ from (C1). From eq. (23), we have $B + E = -A.$ Substitute this and use EWSs in eq. (27), expand eqs. (C1) to derive:

$$C_{T1} = -A(1 - \theta_{T2})(\theta_2/\theta_K)S + B\lambda_{K2}T + E\lambda_{L2}U. \qquad (32)$$

Similarly, expanding eq. (B12), we derive

$$C_{T2} = -A(1 - \theta_{T1})(\theta_1/\theta_K)S + B\lambda_{K1}T + E\lambda_{L1}U. \qquad (33)$$

$C_{Tj}$ is a linear function in $S$, $T$, and $U$. Using EWS-ratios in eq. (26), transform eq. (32) and (33). Substitute in (30) to derive

$$\begin{bmatrix} X_1^*/w_T^* \\ X_2^*/w_T^* \end{bmatrix} = \frac{1}{\Delta} \begin{bmatrix} -E\lambda_{L2}T[U' - f_{T1}(S')] \\ E\lambda_{L1}T[U' - f_{T2}(S')] \end{bmatrix}, \qquad (34)$$

where

$$f_{T1}(S') = [A(1 - \theta_{T2})(\theta_2/\theta_K)S' - B\lambda_{K2}](E\lambda_{L2})^{-1},$$

$$f_{T2}(S') = [A(1 - \theta_{T1})(\theta_1/\theta_K)S' - B\lambda_{K1}](E\lambda_{L1})^{-1}. \qquad (35)$$

Using the Hadamard product of vectors, transform eq. (34). Its sign pattern is expressed:

$$\text{sgn}\begin{bmatrix} X_1{}^*/w_T{}^* \\ X_2{}^*/w_T{}^* \end{bmatrix} = \text{sgn}\frac{1}{\Delta}\begin{bmatrix} -1 \\ 1 \end{bmatrix} \circ \begin{bmatrix} E\lambda_{L2} \\ E\lambda_{L1} \end{bmatrix} \circ T\begin{bmatrix} U' - f_{T1}(S') \\ U' - f_{T2}(S') \end{bmatrix}. \tag{36}$$

In general, if $\mathbf{A} = [a_{ij}]$ and $\mathbf{B} = [b_{ij}]$ are each m x n matrices, their Hadamard product is the matrix of element-wise products, that is, $\mathbf{A} \circ \mathbf{B} = [a_{ij}b_{ij}]$. For this definition, see, for example, Styan (1973, p. 217-18). The Hadamard product is known, for example, in the literature of statistics.

From eq. (34), we derive

$$X_j{}^*/w_T{}^* = 0 \leftrightarrow U' = f_{Tj}(S'), j = 1, 2. \tag{37}$$

This equation expresses the straight line in two dimensions. We call it the equation for line *Tj*, which expresses the border line for a sign pattern of $X_j{}^*/w_T{}^*$ to change.

Using eqs. (37) and (28), make a system of equations. From this, we obtain a quadratic equation in *S'* for each *j*. Solve this to derive two solutions. Each solution denotes the *S'* coordinate value of the intersection point of line *Tj* and the EWS-ratio vector boundary. The solutions are for lines T1 and T2:

$$S' = B/A, -\theta_{K2}/(1-\theta_{T2}), S' = B/A, -\theta_{K1}/(1-\theta_{T1}). \tag{38}$$

In summary, there are three intersection points. Each line *Tj* passes through the same point, which we call point *Q*. The Cartesian coordinates of point *Q* are

$$(S', U') = (\frac{B}{A}, \frac{B}{E}\frac{\theta_L}{\theta_K}). \tag{39}$$

We call two intersection points other than point *Q*, the point $R_{Tj}, j = 1, 2$. The Cartesian coordinates of points $R_{T1}$ and $R_{T2}$ are:

$$(S', U') = (\frac{-\theta_{K2}}{1-\theta_{T2}}, \frac{\theta_{K2}}{\theta_{L2}}\frac{\theta_L}{\theta_K}), (\frac{-\theta_{K1}}{1-\theta_{T1}}, \frac{\theta_{K1}}{\theta_{L1}}\frac{\theta_L}{\theta_K}). \tag{40}$$

Substituting eq. (24) in eq. (39), we derive the sign pattern of point *Q*, that is,

$$\text{sgn}(S', U') = (-, -). \tag{41}$$

This implies that point *Q* belongs to quadrant IV.

The sign patterns of points $R_{T1}$ and $R_{T2}$ are

$$\text{sgn}(S',U') = (-,+),(-,+). \tag{42}$$

Hence, both of these points are in quadrant II. Compare U' values in eq. (40). Using eq. (20), we can easily show that point $R_{T1}$ is above $R_{T2}$.

From eqs. (39)-(42), we can draw points $Q$ and $R_{Tj}$ and hence, line $Tj$ in the figure. Each line $Tj$ divides the region of the EWS-ratio vector into six subregions, that is, subregion P1-3 and M1-3 (see Fig. 1). The sign patterns of vector $[U' - f_{Tj}(S')]$ for each subregion are

$$\text{sgn}\begin{bmatrix} U' - f_{T1}(S') \\ U' - f_{T2}(S') \end{bmatrix} = \begin{matrix} P1 & P2 & P3 & M1 & M2 & M3 \\ \begin{bmatrix} - \\ - \end{bmatrix}, & \begin{bmatrix} - \\ + \end{bmatrix}, & \begin{bmatrix} + \\ + \end{bmatrix}, & \begin{bmatrix} + \\ + \end{bmatrix}, & \begin{bmatrix} + \\ - \end{bmatrix}, & \begin{bmatrix} - \\ - \end{bmatrix} \end{matrix}. \tag{43}$$

Here, from eq. (29) and Fig. 1, the sign of $T \ (= g_{LT})$ for each subregion are

$$\begin{matrix} P1 & P2 & P3 & M1 & M2 & M3 \\ T = (+), & (+), & (+), & (-), & (-), & (-). \end{matrix} \tag{44}$$

Recall that $\Delta < 0$ (see eq. (B2)) and we assume $E > 0$ (see eq. (24)). Substituting these, eqs (43) and (44) in eq. (36), we can state as follows. The sign patterns of vector $[X_j^*/w_T^*]$ for subregions P1–3 and M1-3 are

$$\text{sgn}\begin{bmatrix} X_1^*/w_T^* \\ X_2^*/w_T^* \end{bmatrix} = \begin{matrix} P1 & P2 & P3 & M1 & M2 & M3 \\ \begin{bmatrix} - \\ + \end{bmatrix}, & \begin{bmatrix} - \\ - \end{bmatrix}, & \begin{bmatrix} + \\ - \end{bmatrix}, & \begin{bmatrix} - \\ + \end{bmatrix}, & \begin{bmatrix} - \\ - \end{bmatrix}, & \begin{bmatrix} + \\ - \end{bmatrix} \end{matrix}. \tag{45}$$

There are six patterns in total. Therefore, we make the following statements.

(i) If the EWS-ratio vector $(S',U')$ exists in subregion P1 or M1, the effects of energy price on commodity output in sector 1 and sector 2 are negative and positive, respectively.

(ii) If the EWS-ratio vector exists in subregion P2 or M2, the effects of energy price on commodity output in both sectors 1 and 2 are negative.

(iii) If the EWS-ratio vector exists in subregion P3 or M3, the effects of energy price on commodity output in sector 1 and sector 2 are positive and negative, respectively.

Here, we assume $(S',U') = (+,-)$. The EWS-ratio vector exists in quadrant IV. This implies that energy and capital, extreme factors, are economy-wide complements (see eq. (29)). The sign pattern for subregion P1 in eq. (45) holds. The following result has been established.

*Theorem 1.* We assume that sector 1 is relatively energy-intensive, sector 2 is relatively capital-intensive, labor is the middle factor, and energy and capital are extreme factors. Further, we assume that the middle factor is used relatively intensively in sector 1. Furthermore, if the EWS-ratio vector (*S', U'*) exists in quadrant IV (or subregion P1), in other words, if energy and capital, extreme factors, are economy-wide complements, the effects of energy price on commodity output in sector 1 and sector 2 are negative and positive, respectively.

2.6. Commodity price–commodity output relationship

We analyze the commodity price–commodity output relationship. Using a similar method to expand $C_{T1}$ in eq. (B8) (see eq. (C1)), expand $C_{21}$ in eq. (B6) and use EWSs in eq. (27). We derive

$$C_{21} = aS + bT + c\,U, \tag{46}$$

where

$$a = (\theta_{K1} + \theta_{L1})(\theta_{K2} + \theta_{L2})(\theta_2/\theta_K), b = \theta_{K1}\lambda_{K2}, c = \theta_{L1}\lambda_{L2}.$$

$C_{21}$ is a linear function in *S, T*, and *U*. Using EWS-ratios in eq. (26), transform eq. (46). Substitute in (31) to derive:

$$X_1^*/p_2^* = cT[U' - f_{21}(S')]/\Delta, \tag{47}$$

where

$$f_{21}(S') = -(a/c)S' - b/c. \tag{48}$$

The sign of eq. (47) is expressed:

$$\operatorname{sgn} X_1^*/p_2^* = \operatorname{sgn} cT[U' - f_{21}(S')]/\Delta. \tag{49}$$

From eq. (47), we derive

$$X_1^*/p_2^* = 0 \leftrightarrow U' = f_{21}(S'). \tag{50}$$

This equation expresses the straight line in two dimensions. We call it the equation for line 21, which expresses the border line for a sign of $X_1*/p_2*$ to change. The gradient and intercept of line 21 are $-a/c(<0)$ and $-b/c(<0)$.

Using eqs. (50) and (28), make a system of equations. From this, we obtain a quadratic equation in S':

$$\frac{a}{c}S'^2 + (\frac{a}{c}+\frac{b}{c}-\frac{\theta_L}{\theta_K})S' + \frac{b}{c} = 0. \tag{51}$$

Solve this to derive two distinct real solutions (see eq. (D5)). Each solution denotes the S' coordinate value of the intersection point of line 21 and the EWS-ratio vector boundary. The Cartesian coordinates of the point are the same as points $R_{T1}$ and $R_{T2}$ (see eq. (40)).

Hence, we can draw line 21 in the figure (see Fig. 2). Line 21 divides the region of the EWS-ratio vector into three subregions: subregion Pa-b and Ma. The sign of $U'-f_{21}(S')$ for each subregion are

$$\text{Pa} \quad \text{Pb} \quad \text{Ma}$$
$$\text{sgn}(U'-f_{21}(S')) = (+), (-), (-). \tag{52}$$

Here, from eq. (29) and Fig. 2, the sign of $T \, (= g_{LT})$ for each subregion are

$$\text{Pa} \quad \text{Pb} \quad \text{Ma}$$
$$T = (+), (+), (-). \tag{53}$$

Recall that $\Delta < 0.$ (see eq. (B2)). Substituting this, eqs. (52), and (53) in (49), the sign for each subregion are

$$\text{Pa} \quad \text{Pb} \quad \text{Ma}$$
$$\text{sgn } X_1*/p_2* = (-), (+), (-). \tag{54}$$

From eq. (54), we make the following statements.

(i) If the EWS-ratio vector (S', U') exists in subregion Pa or Ma, the effect of the price of commodity 2 on the output of commodity 1 is negative.

(ii) If the EWS-ratio vector exists in subregion Pb, the effect of the price of commodity 2 on the output of commodity 1 is positive.

(ii) is against Thompson's (1983, p. 48) statement that if the price of good 1 rises, the output of good

2 falls.

Factor-intensity ranking is important only to decide the relative position of points $R_{T1}$ and $R_{T2}$ (compare eqs (20) and (40)). Hence, without regard to the factor-intensity ranking, subregion Pb exits in quadrant II, in which capital and labor are economy-wide complements. Hence, the following result has been established.

*Theorem 2* If the EWS-ratio vector exists in quadrant II, in other words, capital and labor, domestic factors, are economy-wide complements, the effect of the price of commodity 2 on the output of commodity 1 can be positive. This holds without regard to the factor- intensity ranking.

Let industries 1 and 2, respectively, signify exportables and importables. We may conclude as follows. If capital and labor, domestic factors, are economy-wide complements, the strengthening of an import restriction, which raises the domestic price of importables, can increase the commodity output of exportables. Similarly, the reduction of import restrictions, which decreases the domestic price of importables, can decrease the commodity output of exportables.

Using eqs (B6) and (31), we can show that

$$\text{sgn } X_1^*/p_1^* = (+). \tag{55}$$

I omit the proof owing to space constraints. Hence, if the price of good 1 rises, the output of good 1 also rises. This is not against Thompson's (1983, p. 48) statement that if the price of good 1 increases, the output of good 1 also increases.

2.7. Factor endowment–commodity output relationship

From eqs (B9), (B6), (B10) and (B11), we derive

$$X_1^*/V_K^* < 0, X_1^*/V_L^* > 0, X_2^*/V_K^* > 0, X_2^*/V_L^* < 0. \tag{56}$$

If capital increases, output in sector 1 decreases and that in sector 2 increases. If labor increases, output in sector 1 increases and that in sector 2 decreases. Thompson (1983, eq. (8)) called this relationship Rybczynski result.

3. Conclusion

In this study, we assumed a certain pattern of factor-intensity ranking, including one for the middle factor. We assumed sector 1 is relatively energy-intensive, sector 2 is relatively capital-intensive, labor is the middle factor, and energy and capital are extreme factors. Further, we assumed the middle factor is used relatively intensively in sector 1.

At the outset of this study, we posed the following questions.

(i) Can we derive a sufficient condition for each sign pattern of the energy price–commodity output relationship to hold?

(ii) Is Thompson's (1983) statement—if the price of good 1 increases, the output of good 1 rises and that of good 2 falls— plausible?

We derived the following results.

Answer to (i): We analyzed the energy price–commodity output relationship using the EWS-ratio vector. The EWS-ratio vector boundary demarcates the boundary of the region where the EWS ratio vector can exist. Line $Tj$ divides this region into six subregions. There are six patterns of the energy price–commodity output relationship. We derived a sufficient condition for each sign pattern of the relationship to hold. That is, the position of the EWS-ratio vector determines the relationship. Notably, if the EWS-ratio vector ($S',U'$) exists in quadrant IV, in other words, if energy and capital, extreme factors, are economy-wide complements, a specific result holds necessarily (see Theorem 1).

Answer to (ii): We analyzed the commodity price–commodity output relationship. Line 21 divides the region where the EWS-ratio vector can exist into three subregions. Notably, if the EWS-ratio vector exists in quadrant II, in other words, capital and labor, domestic factors, are economy-wide complements, the effect of the price of commodity 2 on the output of commodity 1 can be positive (see Theorem 2). In other words, the cross-price effect on output can be positive. In addition, the effect can be observed without regard to the factor-intensity ranking. This is against Thompson's (1983, p. 48) statement that if the price of good 1 rises, the output of good 2 falls. In general, this can occur in a three-factor, two-good neoclassical model with one factor payment being exogenous (e.g., energy tariff, carbon tax, or greenhouse gas emission traded internationally). This suggests as follows. The strengthening of an import restriction, which raises the domestic price of importables, can increase the commodity output of exportables. Similarly, the reduction of import restrictions, which decreases the domestic price of importables, can decrease the commodity output of exportables. This seems paradoxical.

On the other hand, factor endowment–commodity output relationship is the same as in Thompson (1983, eq. (8)) (see eq. (56)).

In general, we expect that the positive cross-price effect on output will occur in an N-factor, two-good model, where some factor payments are exogenous, if complementarity exists. The findings of this study can be further expanded by researchers conducting simulation studies using a computable general equilibrium model.

Equation Section (Next)

Appendix A: Comment on Thompson (2014)

We show that Thompson's (2014) equations include errors by comparing the equations with ours. We analyze the commodity price–energy import relationship. We define national income as follows (see also footnote 5).

$$I' = I - w_T V_T = p_1 X_1 + p_2 X_2 - w_T V_T. \tag{A1}$$

$I$ is gross national income (see eq. (15)). We use a method similar to that in BC (p. 36), who used the reciprocity relations derived by Samuelson. Partially differentiate (A1) to derive:

$$\frac{\partial I'}{\partial p_j} = X_j, \ \frac{\partial I'}{\partial w_T} = -V_T. \tag{A2}$$

It follows that

$$\frac{\partial}{\partial p_j}(-V_T) = \frac{\partial}{\partial p_j}\frac{\partial I'}{\partial w_T} = \frac{\partial}{\partial w_T}\frac{\partial I'}{\partial p_j} = \frac{\partial X_j}{\partial w_T} \leftrightarrow \frac{V_T{}^*}{p_j{}^*} = -\frac{\theta_j}{\theta_T}\frac{X_j{}^*}{w_T{}^*}, \ j=1,2. \tag{A3}$$

Hence, the commodity price–energy import relationship is a dual counterpart in the energy price–commodity output relationship. Equation (A3) is equivalent to eq. (11) in Thompson (1983). However, he did not present supporting proof.

Substituting eqs (32) in (30), we derive

$$X_1{}^*/w_T{}^* = -\{-A(1-\theta_{T2})(\theta_2/\theta_K)S + B\lambda_{K2}T + E\lambda_{L2}U\}/\Delta. \tag{A4}$$

Substituting eq. (A4) in (A3), we derive

$$V_T{}^*/p_1{}^* = (\theta_1/\theta_T)\{-A(1-\theta_{T2})(\theta_2/\theta_K)S + B\lambda_{K2}T + E\lambda_{L2}U\}/\Delta. \tag{A5}$$

However, eq. (A5) is not equivalent to eq. (11) in Thompson (2014). See below.

Thompson (2014) defined as follows (see eqs (5) and (9) in Thompson (2014)).

$$\sigma_{KE} \equiv \sum_j \lambda_{Kj}(a_{Kj}'/\tau'), \ \lambda_{EK} \equiv \lambda_{E1}\lambda_{K2} - \lambda_{E2}\lambda_{K1}. \tag{A6}$$

Prime ' denotes a percentage change. $\sigma_{KE}$ is an example of the elasticities of input substitution (or EWS in our terminology). E is energy, K capital, and L labor. Thompson (2014) called $\lambda_{ij}$ the industry employment shares. Thompson's definition of $\lambda_{ij}$ is similar to that in the present study.

Equation (11) in Thompson (2014) expresses the effects of prices on energy imports ($E$):

$$E'/p_1' = (\theta_{K2}\sigma_1 - \theta_{L2}\sigma_2)/\Delta, \tag{A7}$$

where

$$\sigma_1 \equiv (\lambda_{KL} - \lambda_{EK})\sigma_{EL} - (\lambda_{EL} + \lambda_{EK})\sigma_{KL}, \sigma_2 \equiv (\lambda_{KL} + \lambda_{EL})\sigma_{EK} + (\lambda_{EK} + \lambda_{EL})\sigma_{LK}. \tag{A8}$$

Replacing $E$ and $\sigma_{ih}$ in eq. (A8) with $T$ and $g_{ih}$, we derive

$$\sigma_1 \equiv (\lambda_{KL} - \lambda_{TK})g_{TL} - (\lambda_{TL} + \lambda_{TK})g_{KL}, \sigma_2 \equiv (\lambda_{KL} + \lambda_{TL})g_{TK} + (\lambda_{TK} + \lambda_{TL})g_{LK}. \tag{A9}$$

Substitute eq. (A9) in (A7). Using eq. (15), eliminate $g_{KL}, g_{TL}, g_{TK}$ and use $g_{LK}, g_{LT}, g_{KT}$:

$$E'/p_1' = \frac{1}{\Delta}\begin{Bmatrix} -\theta_{K2}(\lambda_{TL} + \lambda_{TK})g_{LK}(\theta_L/\theta_K) - \theta_{L2}(\lambda_{TK} + \lambda_{TL})g_{LK} \\ +\theta_{K2}(\lambda_{KL} - \lambda_{TK})g_{LT}(\theta_L/\theta_T) - \theta_{L2}(\lambda_{KL} + \lambda_{TL})g_{KT}(\theta_K/\theta_T) \end{Bmatrix}. \tag{A10}$$

For example, from eq. (A6), we can show that

$$\lambda_{KL} + \lambda_{TL} = \lambda_{K1}\lambda_{L2} - \lambda_{K2}\lambda_{L1} + \lambda_{T1}\lambda_{L2} - \lambda_{T2}\lambda_{L1} = (\lambda_{K1} - \lambda_{L1}) + (\lambda_{T1} - \lambda_{L1}). \tag{A11}$$

Substituting eq. (A11) in (A10), we have for the term of $g_{KT}$:

$$\{-\theta_{L2}[(\lambda_{K1} - \lambda_{L1}) + (\lambda_{T1} - \lambda_{L1})]g_{KT}(\theta_K/\theta_T)\}/\Delta. \tag{A12}$$

Equation (A10) must be equivalent to (A5). Hence, eq. (A12) must be equivalent to the term of $U (= g_{KT})$ in eq. (A5), that is,

$$(\theta_1/\theta_T)E\lambda_{L2}U/\Delta. \tag{A13}$$

However, this does not hold. Similarly, we can show that other terms in eq. (A10) also include errors.

Appendix B: Solution of eq. (19)

Using Cramer's rule to solve eq. (19) for $X_1^*$, we derive

$$X_1^* = \Delta_4 / \Delta, \tag{B1}$$

where $\Delta = \det(\mathbf{A})$, $\Delta_4 = \det(\mathbf{A}_4) = \begin{vmatrix} 0 & \theta_{K1} & \theta_{L1} & p_1^* - \theta_{T1}w_T^* & 0 \\ 0 & \theta_{K2} & \theta_{L2} & p_2^* - \theta_{T2}w_T^* & 0 \\ -1 & g_{TK} & g_{TL} & -g_{TT}w_T^* & \lambda_{T2} \\ 0 & g_{KK} & g_{KL} & V_K^* - g_{KT}w_T^* & \lambda_{K2} \\ 0 & g_{LK} & g_{LL} & V_L^* - g_{LT}w_T^* & \lambda_{L2} \end{vmatrix}.$

$\Delta$ is the determinant of matrix $\mathbf{A}$. Replacing column 4 of matrix $\mathbf{A}$ with column vector $\mathbf{P}$, we derive matrix $\mathbf{A}_4$. $\Delta_4$ is the determinant of matrix $\mathbf{A}_4$. Express the above as a cofactor expansion along the first column:

$$\Delta = -\begin{vmatrix} \theta_{K1} & \theta_{L1} \\ \theta_{K2} & \theta_{L2} \end{vmatrix} \begin{vmatrix} \lambda_{K1} & \lambda_{K2} \\ \lambda_{L1} & \lambda_{L2} \end{vmatrix} = -(\theta_{K1}\theta_{L2} - \theta_{K2}\theta_{L1})^2 \frac{\theta_1 \theta_2}{\theta_K \theta_L} < 0, \tag{B2}$$

$$\Delta_4 = -[|J| - w_T^*|K|], \tag{B3}$$

where

$$|J| = \begin{vmatrix} \theta_{K1} & \theta_{L1} & p_1^* & 0 \\ \theta_{K2} & \theta_{L2} & p_2^* & 0 \\ g_{KK} & g_{KL} & V_K^* & \lambda_{K2} \\ g_{LK} & g_{LL} & V_L^* & \lambda_{L2} \end{vmatrix}, |K| = \begin{vmatrix} \theta_{K1} & \theta_{L1} & \theta_{T1} & 0 \\ \theta_{K2} & \theta_{L2} & \theta_{T2} & 0 \\ g_{KK} & g_{KL} & g_{KT} & \lambda_{K2} \\ g_{LK} & g_{LL} & g_{LT} & \lambda_{L2} \end{vmatrix}. \tag{B4}$$

Express the above as a cofactor expansion along the third column:

$$|J| = p_1^* C_{11} + p_2^*(-C_{21}) + V_K^* C_{K1} + V_L^*(-C_{L1}), \tag{B5}$$

where

$$C_{11} = \begin{vmatrix} \theta_{K2} & \theta_{L2} & 0 \\ g_{KK} & g_{KL} & \lambda_{K2} \\ g_{LK} & g_{LL} & \lambda_{L2} \end{vmatrix}, C_{21} = \begin{vmatrix} \theta_{K1} & \theta_{L1} & 0 \\ g_{KK} & g_{KL} & \lambda_{K2} \\ g_{LK} & g_{LL} & \lambda_{L2} \end{vmatrix}, C_{K1} = \begin{vmatrix} \theta_{K1} & \theta_{L1} & 0 \\ \theta_{K2} & \theta_{L2} & 0 \\ g_{LK} & g_{LL} & \lambda_{L2} \end{vmatrix}, C_{L1} = \begin{vmatrix} \theta_{K1} & \theta_{L1} & 0 \\ \theta_{K2} & \theta_{L2} & 0 \\ g_{KK} & g_{KL} & \lambda_{K2} \end{vmatrix}.$$

(B6)

Next, sum up columns 1 and 2 in column 3 and subtract row 2 from row 1. Next, express as a cofactor expansion along the third column. We have

$$|K| = \begin{vmatrix} B & E & 0 & 0 \\ \theta_{K2} & \theta_{L2} & 1 & 0 \\ g_{KK} & g_{KL} & 0 & \lambda_{K2} \\ g_{LK} & g_{LL} & 0 & \lambda_{L2} \end{vmatrix} = -C_{T1}, \tag{B7}$$

where recall eq. (22), that is, $(A, B, E) = (\theta_{T1} - \theta_{T2}, \theta_{K1} - \theta_{K2}, \theta_{L1} - \theta_{L2})$ and where

$$C_{T1} = \begin{vmatrix} B & E & 0 \\ g_{KK} & g_{KL} & \lambda_{K2} \\ g_{LK} & g_{LL} & \lambda_{L2} \end{vmatrix}. \tag{B8}$$

From eqs. (B1), (B3), (B5), and (B7), we have

$$X_1^* = -\left[ p_1^* C_{11} + p_2^* (-C_{21}) + V_K^* C_{K1} + V_L^* (-C_{L1}) + w_T^* C_{T1} \right] / \Delta. \tag{B9}$$

Similarly, solve eq. (19) for $X_2^*$ to derive:

$$X_2^* = -\left[ p_1^* (-C_{12}) + p_2^* C_{22} + V_K^* (-C_{K2}) + V_L^* C_{L2} - w_T^* C_{T2} \right] / \Delta, \tag{B10}$$

where

$$C_{12} = \begin{vmatrix} \theta_{K2} & \theta_{L2} & 0 \\ g_{KK} & g_{KL} & \lambda_{K1} \\ g_{LK} & g_{LL} & \lambda_{L1} \end{vmatrix}, C_{22} = \begin{vmatrix} \theta_{K1} & \theta_{L1} & 0 \\ g_{KK} & g_{KL} & \lambda_{K1} \\ g_{LK} & g_{LL} & \lambda_{L1} \end{vmatrix}, C_{K2} = \begin{vmatrix} \theta_{K1} & \theta_{L1} & 0 \\ \theta_{K2} & \theta_{L2} & 0 \\ g_{LK} & g_{LL} & \lambda_{L1} \end{vmatrix}, C_{L2} = \begin{vmatrix} \theta_{K1} & \theta_{L1} & 0 \\ \theta_{K2} & \theta_{L2} & 0 \\ g_{KK} & g_{KL} & \lambda_{K1} \end{vmatrix},$$

(B11)

$$C_{T2} = \begin{vmatrix} B & E & 0 \\ g_{KK} & g_{KL} & \lambda_{K1} \\ g_{LK} & g_{LL} & \lambda_{L1} \end{vmatrix}. \tag{B12}$$

Equation Section (Next)

Appendix C: Expansion of eq. (B8)

Using eq. (17), expand eq. (B8) to have

$$C_{T1} = \begin{vmatrix} B & E & 0 \\ -g_{KL} & g_{KL} & \lambda_{K2} \\ g_{LK} & g_{LL} & \lambda_{L2} \end{vmatrix} + \begin{vmatrix} 0 & E & 0 \\ -g_{KT} & g_{KL} & \lambda_{K2} \\ 0 & g_{LL} & \lambda_{L2} \end{vmatrix}$$

$$= \begin{vmatrix} B & E & 0 \\ -g_{KL} & g_{KL} & \lambda_{K2} \\ g_{LK} & -g_{LK} & \lambda_{L2} \end{vmatrix} + \begin{vmatrix} B & 0 & 0 \\ -g_{KL} & 0 & \lambda_{K2} \\ g_{LK} & -g_{LT} & \lambda_{L2} \end{vmatrix} - E(-g_{KT})\lambda_{L2}$$

$$= -\lambda_{K2}\begin{vmatrix} B & E \\ g_{LK} & -g_{LK} \end{vmatrix} + \lambda_{L2}\begin{vmatrix} B & E \\ -g_{KL} & g_{KL} \end{vmatrix} + Bg_{LT}\lambda_{K2} + Eg_{KT}\lambda_{L2}$$

$$= \lambda_{K2}(B+E)g_{LK} + \lambda_{L2}(B+E)g_{KL} + B\lambda_{K2}g_{LT} + E\lambda_{L2}g_{KT}. \tag{C1}$$

Equation Section (Next)

Appendix D: Solutions of eq. (51)

We define for ease of notation,

$$x = \theta_{K1}/\theta_{L1},\ y = \theta_L/\theta_K,\ z = \theta_{K2}/\theta_{L2}. \tag{D1}$$

We derive

$$\frac{a}{c} = (1+x)(1+z)y,\ \frac{b}{c} = xzy,\ \frac{a}{c} + \frac{b}{c} - \frac{\theta_L}{\theta_K} = [(x+z)+2xz]y. \tag{D2}$$

Substitute eq. (D2) in (51) and divide the both sides by $y$. We derive:

$$(1+x)(1+z)S'^2 + [(x+z)+2xz]S' + xz = 0. \tag{D3}$$

The discriminant of eq. (D3) is

$$D = [(x+z)+2xz]^2 - 4(1+x)(1+z)xz = (x-z)^2\ (>0). \tag{D4}$$

Using eq. (20), we derive $D > 0$. Hence, eq. (D3) has two distinct real solutions. The solutions are:

$$S' = -z/(1+z), -x/(1+x) = -\theta_{K2}/(1-\theta_{T2}), -\theta_{K1}/(1-\theta_{T1}). \tag{D5}$$

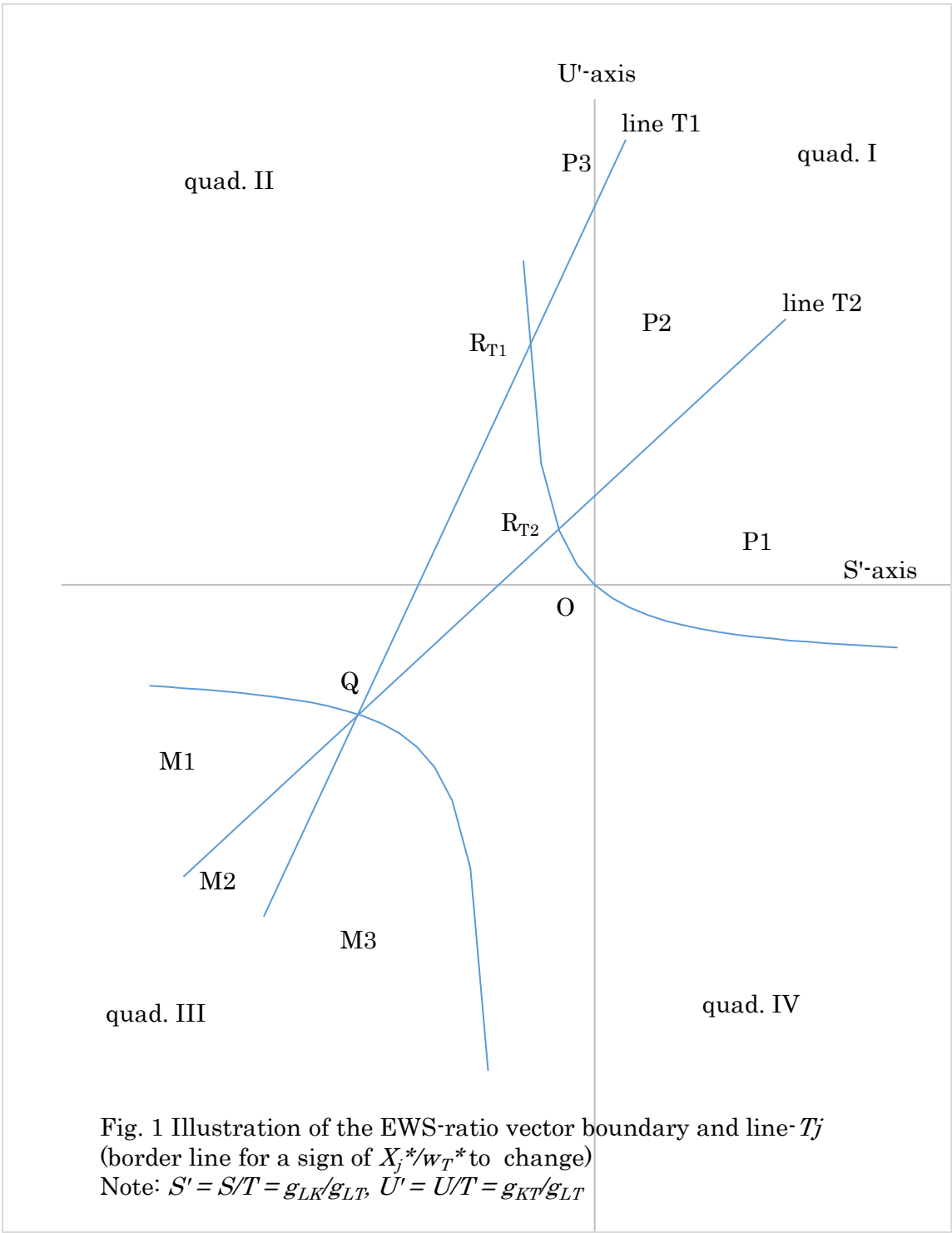

Fig. 1 Illustration of the EWS-ratio vector boundary and line-$Tj$
(border line for a sign of $X_j^*/w_T^*$ to change)
Note: $S' = S/T = g_{LK}/g_{LT}$, $U' = U/T = g_{KT}/g_{LT}$

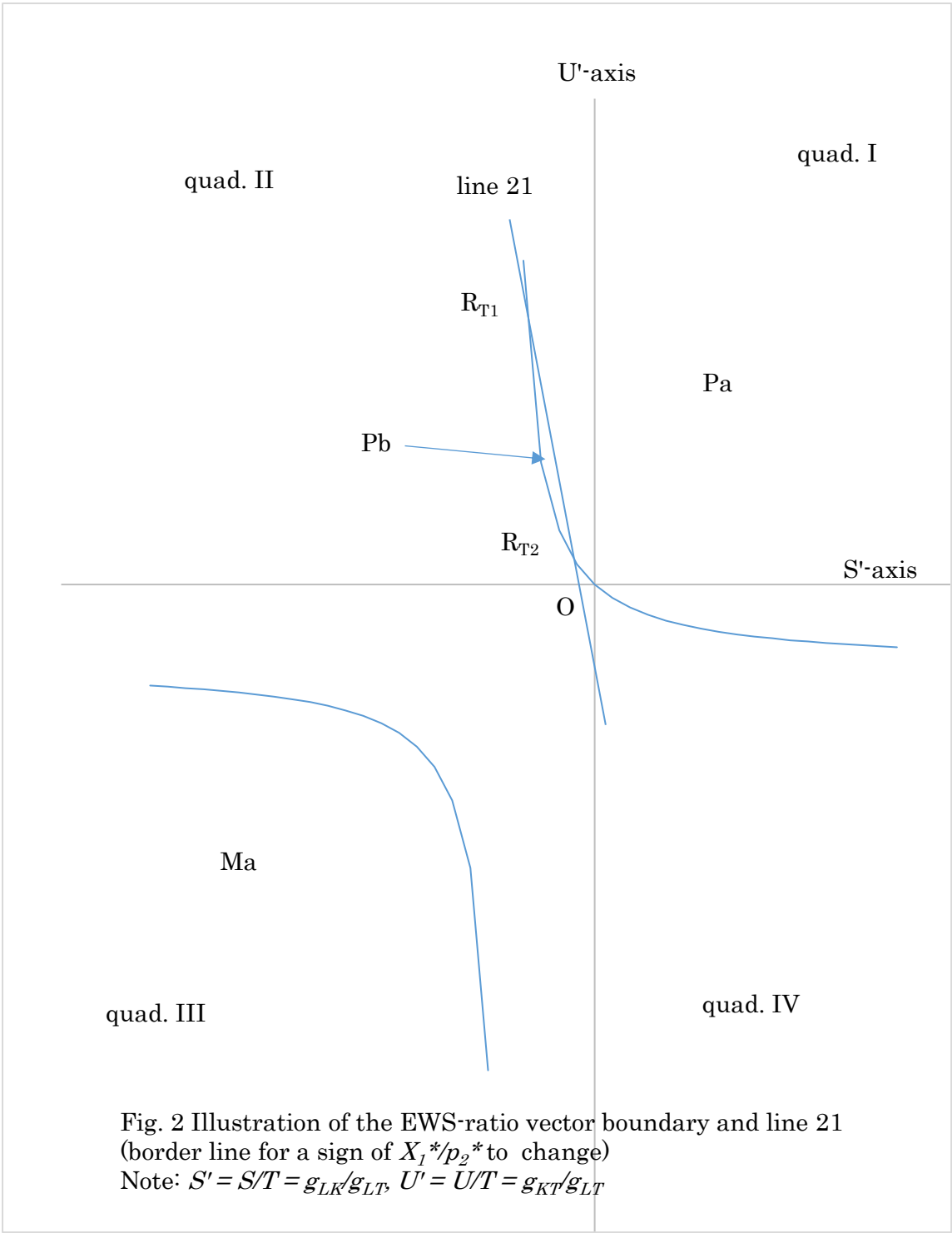

Fig. 2 Illustration of the EWS-ratio vector boundary and line 21 (border line for a sign of $X_1^*/p_2^*$ to change)
Note: $S' = S/T = g_{LK}/g_{LT}$, $U' = U/T = g_{KT}/g_{LT}$